\documentclass{osa-article}

\journal{boe}
\usepackage[section]{placeins}
\usepackage[labelfont=bf,textfont=bf]{caption}
\usepackage{bm}

\begin{document}

\title{Large dynamic range autorefraction with a low-cost diffuser wavefront sensor}

\author{Gregory N. McKay,\authormark{1} Faisal Mahmood,\authormark{1} and Nicholas J. Durr\authormark{1,*}}

\address{\authormark{1}Department of Biomedical Engineering, Johns Hopkins University (JHU), 3400 N. Charles Street, Baltimore, MD 21218, USA}


\email{\authormark{*}ndurr@jhu.edu} 

\homepage{http://durr.jhu.edu} 


\begin{abstract}
 Wavefront sensing with a thin diffuser has emerged as a potential low-cost alternative to a lenslet array for aberrometry. Diffuser wavefront sensors (DWS) have previously relied on tracking speckle displacement and consequently require coherent illumination. Here we show that displacement of caustic patterns can be tracked for estimating wavefront gradient, enabling the use of incoherent light sources and large dynamic-range wavefront measurements. We compare the precision of a DWS to a Shack-Hartmann wavefront sensor (SHWS) when using coherent, partially coherent, and incoherent illumination, in the application of autorefraction. We induce spherical and cylindrical errors in a model eye and use a multi-level Demon's non-rigid registration algorithm to estimate caustic displacements relative to an emmetropic model eye. When compared to spherical error measurements with the SHWS using partially coherent illumination, the DWS demonstrates a $\sim$5-fold improvement in dynamic range (-4.0 to +4.5 D vs. -22.0 to +19.5 D) with less than half the reduction in resolution (0.072 vs. 0.116 D), enabling a $\sim$3-fold increase in the number of resolvable prescriptions (118 vs. 358). In addition to being 40x lower-cost, the unique, non-periodic nature of the caustic pattern formed by a diffuser enables a larger dynamic range of aberration measurements compared to a lenslet array....
\end{abstract}

\section{Introduction}

Visual impairment from uncorrected refractive error affects over a billion people worldwide, and accounts for an estimated \$200 billion in lost productivity annually \cite{Fricke2012,Durr2014,Naidoo2012,Resnikoff2008}. The World Health Organization estimates that the potential gain in productivity from addressing this global health burden could be an order of magnitude greater than the direct cost of conducting such a large-scale intervention  \cite{Smith2009}. Despite growing global health initiatives, significant obstacles remain. One critical impediment is a lack of eye care providers in low-resource settings \cite{Dandona2001}. For example, parts of Sub-Saharan Africa have as few as one ophthalmologist per million people, while more developed nations have on average 80-fold more per capita \cite{Resnikoff2012, WHO2010, Dandona2001}. This disparity has spurred innovation in developing technologies that increase the efficiency, reduce training requirements, and improve quality of care of current providers \cite{Durr2014}. In particular, there are several recent efforts to introduce low-cost, portable autorefractors to improve access to refractive eye care \cite{Durr2015,Ciuffreda2015,Pamplona2010}, and a large field study has shown that accurate prescriptions can be obtained in low-resource settings with a hand-held wavefront aberrometer \cite{Durr2018}.

Many low-cost, portable autorefractors utilize a Shack-Hartmann wavefront sensor (SHWS), which consists of a microlens array placed one focal length in front of an optical sensor. The incident portion of an electromagnetic wavefront scattered off a patient's retina can be sampled by this equally-spaced, periodic lenslet array, and focused onto the plane of the optical sensor to create a spotfield intensity distribution. The location and displacement of each of these spots relative to a reference spotfield allows measurement of the local wavefront curvature, which can be used to generate an eyeglass prescription \cite{Liang1994, Applegate2014, Bruce2014, Cheng2003, Dai2008, Porter2006}. The use of a periodic lenslet array allows for simple, rapid spotfield tracking by using centroiding algorithms to measure spot location and calculate displacement. Beyond their ophthalmic applications, SHWSs are also broadly utilized in metrology, astronomy, and microscopy \cite{Platt2001}. Despite undergoing significant development in software algorithms, microfabrication, and cost reduction over the last several decades, the basic operating principles of the SHWS remain the same, and the microlens array remains a central, high-cost component. Moreover, the limited dynamic range of a microlens array typically requires the addition of a moving Badal lens system to compensate primary aberrations and bring the wavefront from the eye into a measurable range \cite{Wittenberg1988, Atchison1995}. These additional optics and moving parts increase the cost and reduce reliability of wavefront autorefractors.

Optical diffusers have been explored for lensless imaging, lightfield measurements, and imaging through complex media \cite{Huang2103,Antipa2016,Waller2018,Engfei2017}. Wavefront sensing with a thin diffuser has also been demonstrated, relying on the principle of the diffuser memory effect \cite{Erto2017, Kane1988, Freund1988}. Recently, Berto \textit{et al.} demonstrated the fundamentals for using a diffuser to measure high resolution wavefronts using integrated transverse displacement maps, which were generated from non-rigid image registration of speckle pattern distortion between planar and aberrated wavefronts \cite{Erto2017}. Further, Gunjala \textit{et al.} used a statistical approach to reconstruct aberration profiles from multiple images through a weak diffuser at distinct angles of illumination \cite{Gunjala2018}.

\begin{figure}[h!]
\centering\includegraphics[width=12cm]{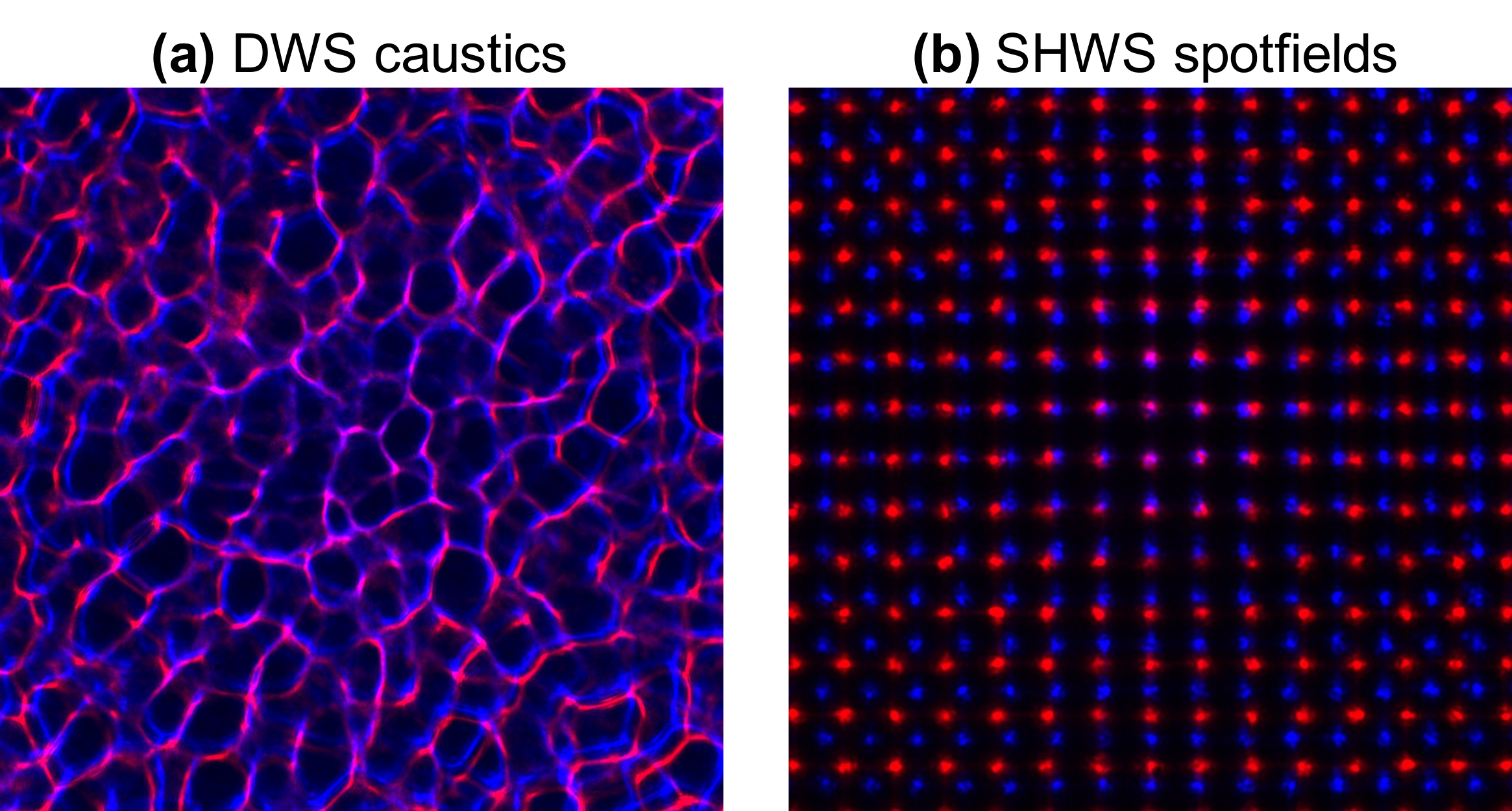}
\caption{(a) Example DWS caustic patterns from an emmetropic model eye (red) and 5D myopic eye (blue), and (b) corresponding spotfields from the SHWS with the same refractive error and pseudocoloring as in (a). Note the periodic nature of the spotfield image, compared to the relatively random caustic pattern of the DWS.}
\label{fig:CausticSpotfield}
\end{figure}

In this report, we develop a diffuser wavefront sensor (DWS) for autorefraction, characterize performance tradeoffs, and compare metrics that are relevant for autorefraction. We show that wavefronts can be measured by the deformation of incoherent caustic intensity patterns, rather than relying on the memory effect of speckle intensity patterns utilized in other works \cite{Erto2017, Gunjala2018, Kane1988, Freund1988}. By constructing and testing a DWS and a SHWS in parallel, we measure refractive error in a model eye and directly compare the performance of the two devices under three different illumination techniques: a laser diode (LD), a laser diode with a laser speckle reducer (LD+LSR), and an incoherent light emitting diode (LED). While we recognize that significant advancements in algorithmic \cite{Leroux2009a, Pfund1998a, Xia2010, Yu2014, Shinto2016} and lenslet microfabrication \cite{Hongbin2008, Aita2015} have been accomplished to increase the dynamic range of a SHWS, we use a conventional SHWS lenslet array and spot tracking algorithm where spots are confined within their lenslet's field of view for robust measurement. We find that a holographic diffuser intrinsically provides an increased dynamic range and number of resolvable prescriptions, while maintaining similar precision and acceptable accuracy when compared to the conventional lenslet array. The performance benefits of the DWS originate from the uniqueness of the caustic pattern produced by the diffuser. Compared to the more uniform spots produced by a lenslet array, unique features from the caustic pattern can be tracked across larger displacement on the sensor without incurring origin ambiguity that arises when a spot in a SHWS translated out of the aperture of a lenslet. Figure~\ref{fig:CausticSpotfield} shows images of caustic patterns acquired by the DWS and spotfields captured by the SHWS.

\captionsetup{labelfont=md,textfont=md}

\section{Theory}

\begin{figure}[h!]
\centering\includegraphics[width=12cm]{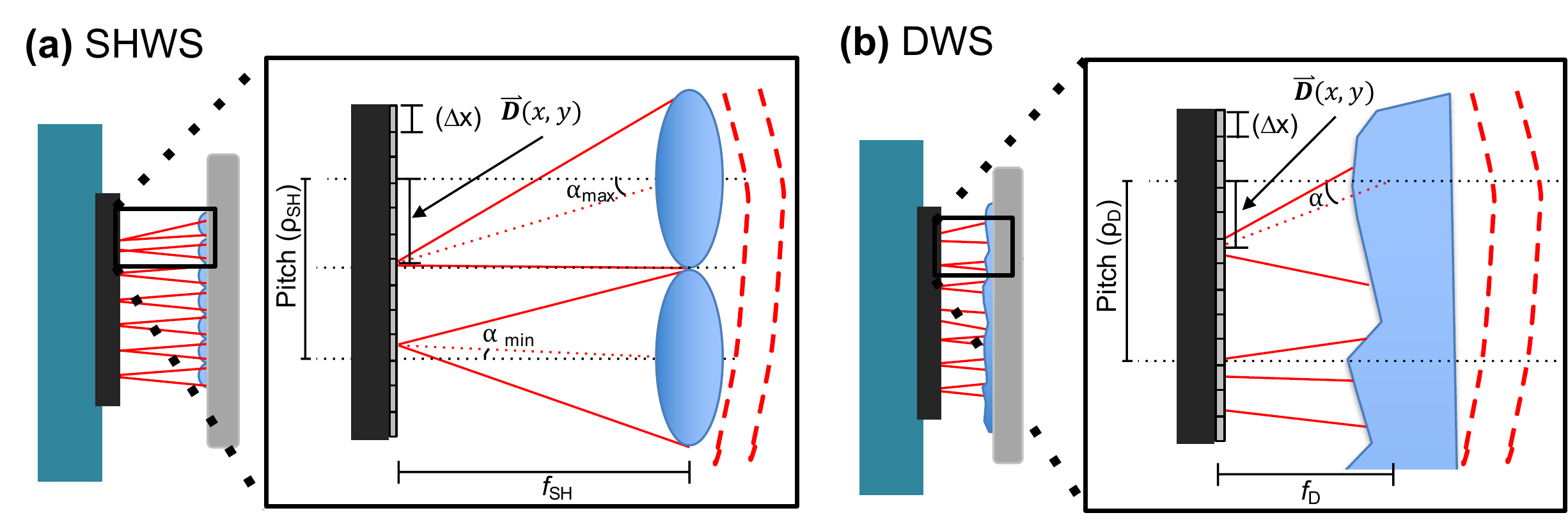}
\caption{(a) Diagram of SHWS fundamental dynamic range and sensitivity constraints. At high wavefront tilts ($\alpha >\alpha_{max}$), spot displacement into an adjacent lenslet's field of view causes origin ambiguity. At low wavefront tilts ($\alpha < \alpha_{min}$), spot displacement is less than a pixel. (b) A similar model is proposed for the holographic diffuser, where the effective diffuser focal length is the distance from diffuser surface to sharp caustic image formation ($f_D$), and the effective diffuser pitch is the measured average distance between sharp caustic intensity peaks ($\rho_D$).}
\label{fig:SHWSDynamicRangeSchematic}
\end{figure}

The dynamic range and sensitivity to the angular tilt of a local wavefront ($\alpha_{max}$ and $\alpha_{min}$) in a SHWS is determined by the sensor pixel size ($\Delta x$), the lenslet pitch ($\rho_{SH}$), and the lenslet focal length ($f_{SH}$), shown schematically in Figure~\ref{fig:SHWSDynamicRangeSchematic}(a). The minimum detectable spot deviation is set by $\Delta x$, while the maximum displacement is set by $\rho_{SH}$/2. The magnitude of spot deviation for a given wavefront tilt scales with the focal length of the lenslets for the $tan(\alpha) \simeq \alpha$ approximation, and thus the angular sensitivity and range are fundamentally opposing constraints in a SHWS design. This can be represented as:

\begin{equation}
\alpha_{max} = \frac{\rho_{SH}}{2 f_{SH}}
\label{eq:SHWSMax}
\end{equation}

\begin{equation}
\alpha_{min} = \frac{\Delta x}{f_{SH}}
\label{eq:SHWSMin}
\end{equation}

\vskip 0.1in

From Equations ~\ref{eq:SHWSMax}-~\ref{eq:SHWSMin}, it is apparent that the number of resolvable tilts, $\alpha_{max} / \alpha_{min}$, is independent of the choice of focal length:

\begin{equation}
\frac{\alpha_{max}}{\alpha_{min}} = \frac{\rho_{SH}}{2  \Delta x }
\label{eq:SHWSRange}
\end{equation}

\vskip 0.1in

Similarly, for an aberrometer, the dynamic range and sensitivity of curvature measurement depends on $\Delta x$, $\rho_{SH}$, and $f_{SH}$ for a given pupil diameter. This tradeoff between sensitivity and range makes the number of resolvable prescriptions a useful figure of merit.

There are no manufacturer's specifications for the 0.5$^{\circ}$ holographic diffuser used in this study that are directly analogous to those of the lenslet array. To compare the dynamic range of the DWS to the SHWS directly, we model the diffuser as a non-periodic lenslet array, shown schematically in Figure~\ref{fig:SHWSDynamicRangeSchematic}(b). We define the effective diffuser pitch ($\rho_D$) as the mean distance between sharp caustic intensity bands, and the effective diffuser focal length ($f_D$) as the distance from the diffuser to the sensor. By taking plot profiles through the caustic pattern produced by the holographic diffuser with on-axis plane wave illumination, we measure $\rho_D$ = 338$\mu$m with a standard deviation of 21$\mu$m. Note that the pitch is not uniform for the DWS, in contrast to the fixed $\rho_{SH}$ = 300$\mu$m of the SHWS. The effective diffuser focal length was empirically chosen to be $f_D$ = 5.15mm based on the distance from the diffuser to the sensor where the caustic pattern appeared to be in sharpest focus.

To compare the DWS and SHWS, one must account for differences in their characteristics. For example, since $\rho_D$ $\approx$ $\rho_{SH}$ and $f_D$ < $f_{SH}$, we would expect the DWS to have an intrinsically lower sensitivity and a higher range than the SHWS. Additionally, differences in pixel size change the sensitivity of the device, and differences in pupil radius over which the sensor measures the wavefront change sensitivity and range. In the case of autorefraction, the further from the optical axis a first-order refractive error is sampled, the greater the local wavefront angle. Thus, as the sampled pupil radius increases, the sensitivity to small refractive error increases, but the dynamic range before spot origin ambiguity occurs decreases. We derive criteria based on a thin lens approximation ray trace to capture the predicted sensitivity and range of the DWS and SHWS in the current system used, shown in Appendix Section \ref{ssec:AppendixDRSensCalc}. Table~\ref{table:PredictedSensitivityDRNumPrescriptions} organizes the relevant characteristics of each device as well as the predicted range, sensitivity, and number of resolvable prescriptions in terms of refractive error. This analysis predicts that the SHWS has a lower dynamic range and increased sensitivity as compared to the DWS. Importantly though, the $\rho/2$ cutoff criteria for measurement range yields a similar predicted $\alpha_{max}$ / $\alpha_{min} $ for both devices, highlighting the tradeoff between sensitivity and dynamic range. Thus, if both devices operate with the conventional $\rho/2$ maximum detectable displacement criteria, they should be able to resolve a similar number of prescriptions.

\captionsetup{labelfont=bf,textfont=bf}
\begin{table}[h!]
\caption{Comparison of Attributes of SHWS and DWS} \label{tab:Measured_DR_MStd_RMSE_title} 
\centering
\begin{tabular}{c c c} 
 \hline
 Wavefront Sensor Attribute & SHWS & DWS \\ [0.5ex]
 \hline
 $f$ (mm) & 14.6 & 5.15 \\
 $\rho$ ($\mu$m) & 300 & 338 $\pm$ 21 \\
 $\Delta x$ ($\mu$m) & 4.65 & 5.2 \\
 Predicted $\alpha_{min}$ (mrad) & 0.318 & 1.000 \\
 Predicted $\alpha_{max}$ (rad) & 0.0103 & 0.0328 \\
 $\alpha_{max}$ / $\alpha_{min} $ & 32.4 & 32.8 \\
 Predicted RE Sensitivity (D) & 0.13 & 0.37 \\
 Predicted RE Dynamic Range (D) & [-4.30, +4.30D] & [-12.30D, +12.30D] \\
 Predicted NRP & 66 & 66 \\
\hline
\end{tabular}
\label{table:PredictedSensitivityDRNumPrescriptions}
\end{table}
\captionsetup{labelfont=md,textfont=md}

\section{Methods}

\subsection{Optical System}

\begin{figure}[h!]
\centering\includegraphics[width=12cm]{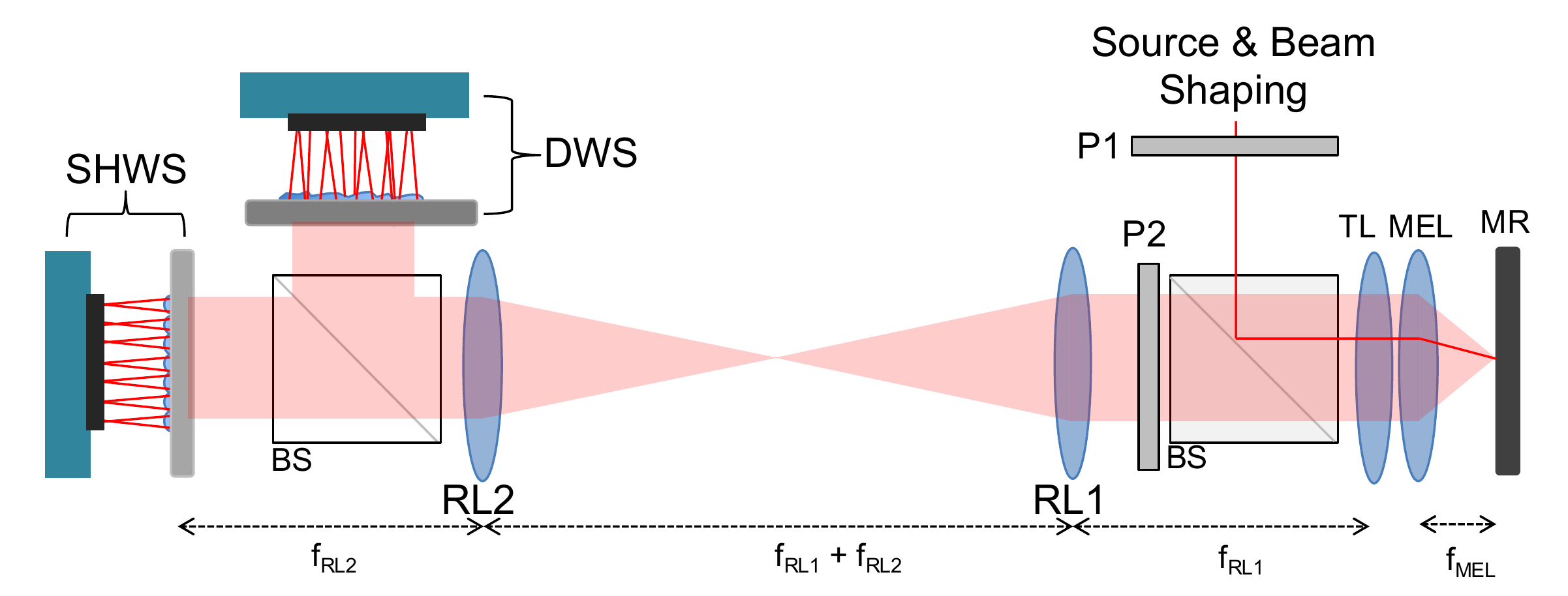}
\caption{Schematic of DWS and SHWS experimental setup. A collimated, slightly off-axis pencil beam was linearly polarized (P1), redirected with a beamsplitter (BS), and focused by the model eye lens (MEL) onto the model retina (MR). The resulting point-source illumination (occurring in the absence of a trial lens (TL)) is re-collimated by the MEL and relayed by a 1x, 4$f$ telescopic system (RL1 and RL2) to the lenslet array of the SHWS and the holographic diffuser of the DWS simultaneously. A second crossed polarizer (P2) is used to mitigate back reflection. Various TLs are introduced, at a conjugate plane to the lenslet and diffuser, to characterize sensitivity and dynamic range.}
\label{fig:ExperimentalSetup}
\end{figure}

Figure~\ref{fig:ExperimentalSetup} shows a schematic of the optical system constructed to measure wavefronts with the DWS and SHWS concurrently. The phase-encoding element of each device (diffuser for DWS and lenslet array for SHWS) was placed conjugate to the trial lens introduced with a 1x, 4$f$ telescopic relay, comprised of two $f$=50mm lenses. The signal to each instrument was separated by a 50:50 beamsplitter.  The SHWS consisted of a Thorlabs WFS300-14AR with lenslet pitch of 300$\mu$m, focal length of 14.6mm, and pixel size 4.65$\mu$m. The DWS consisted of a 0.5$^{\circ}$ holographic diffuser (Edmund Optics) separated from a Thorlabs DCC1545M (pixel size 5.2$\mu$m) by 5.15mm. A holographic diffuser was utilized in the DWS for its sharp caustic pattern formation and high transparency at visible wavelengths. The characteristics of the SHWS and DWS are organized in Table~\ref{table:PredictedSensitivityDRNumPrescriptions}, including the predicted sensitivity and dynamic range in terms of refractive error. The model eye consisted of a biconcave lens ($f$= 25.4mm) separated from a model retina (diffusive piece of paper) by its focal length. Three different illumination systems were tested: (1) laser diode (LD), (2) laser diode + laser speckle reducer (LD+LSR), and (3) light emitting diode (LED), each at 650nm. Each source was shaped to a collimated pencil beam 1mm in diameter and incorporated into the system 2mm off-axis to reduce back reflections. Note that LD+LSR illumination was focused onto the Optotune LSR-3005 before re-collimating, which was accomplished by two $f$=15mm lenses separated by $2f$ \cite{Optotune2014}. The LED was collimated with an $f$=25mm aspheric condenser and an additional $f$=200mm focal length lens. Crossed linear polarizers were incorporated into the system to further reduce back reflection. Trial lenses from a standard optometrist's kit were introduced adjacent to the model eye lens to induce known refractive errors. Trial lens powers with spherical equivalent power (M) from [-24D, +24D] were measured in both the DWS and the SHWS, recording raw images of the distorted caustics with the DWS and measuring centroid positions from the recorded spotfield images of SHWS. Refractive errors exceeding $\pm$12D were achieved with two trial lenses placed close together. In this scenario, the thin lens approximation was decreasingly accurate for higher refractive error, so the changing effective focal length and principal plane location with each trial lens introduced was modeled using a Gaussian reduction, and used to calculate the effective trial lens power measured by the DWS and SHWS. More information about this modeling is shown in Appendix Section \ref{ssec:AppendixEFLBFLPP}. Cylindrical trial lenses from [-4D, +4D] were introduced with their axis at 0$^{\circ}$ and 45$^{\circ}$ to measure Jackson cross-cylinder values ($J_0$ and $J_{45}$) with each wavefront sensor. For each trial lens tested, five measurements were obtained to compare test retest reliability between the DWS and SHWS for each illumination type.

\subsection{Shack-Hartmann Wavefront Sensing Algorithm}

A commercial Shack-Hartmann wavefront sensor (Thorlabs WFS300-14AR) was calibrated to a planar wavefront in the absence of a trial lens (emmetropic condition), and subsequent measurements were taken as each trial lens was introduced to induce refractive error. We analyzed the spotfield centroids calculated by the commercial SHWS. The distances between centroids in the emmetropic and ametropic conditions were calculated, and the resulting x-y displacement map was scaled by the known pixel size and divided by the lenslet focal length ($f_{SH}$ = 14.6mm). The Zernike coefficients were fit to the displacement field data in the first derivative domain using a pseudo-inverse approach \cite{Schwiegerling}. The fitted Zernike coefficients were used to calculate M, $J_0$, and $J_{45}$, as described below for the DWS.

\subsection{Diffuser Wavefront Sensing Algorithm}

\begin{figure}[h!]
\centering\includegraphics[width=12cm]{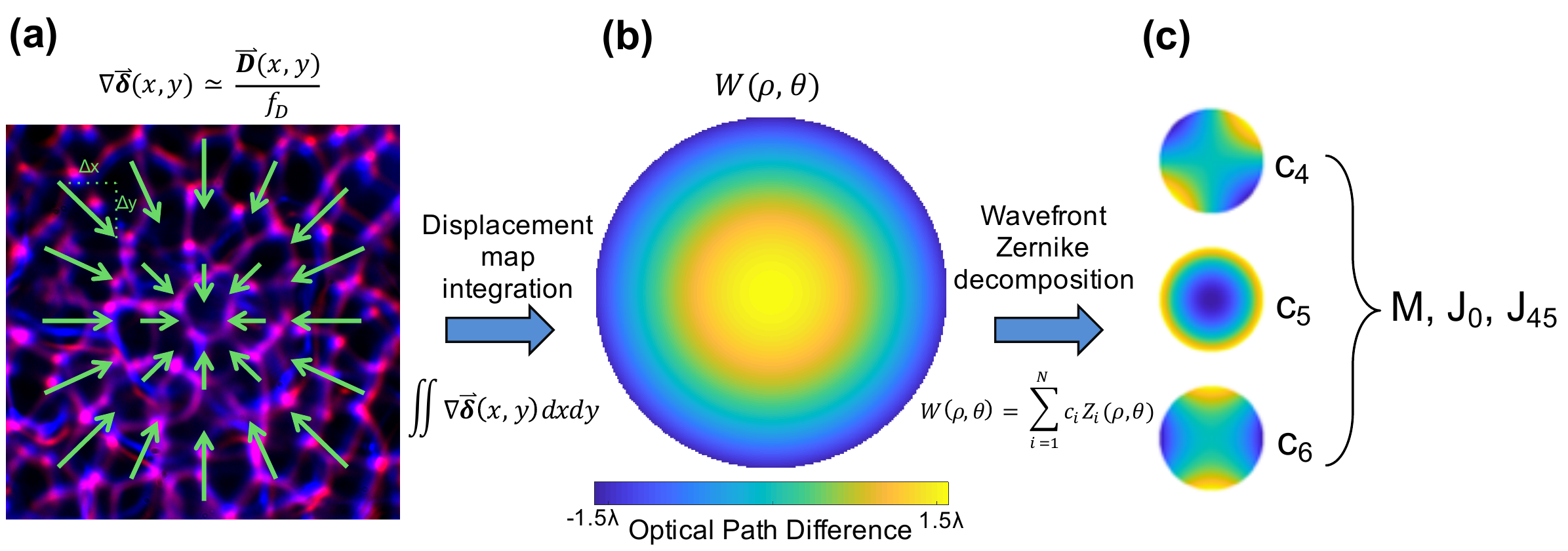}
\caption{Outline of the DWS algorithm. (a) A reference caustic from an emmetropic eye (red) is registered to a distorted caustic from an ametropic eye (blue). The resulting vector field displacement map ($\boldsymbol{\vec{D}}(x,y)$, green arrows), is scaled to produce the transverse gradient of the wavefront ($\nabla \boldsymbol{\vec{\delta}}(x,y)$). (b) 2D gradient integration yields a measurement of the wavefront ($W(\rho, \theta)$). (c) The measured wavefront is decomposed into a Zernike basis and a refractive error is calculated.}
\label{fig:DWSAlgorithm}
\end{figure}

Wavefronts were calculated from caustic pattern distortions captured with a DWS, using a similar approach to Berto \textit{et al.} \cite{Erto2017}. An overview of these steps is shown in Figure~\ref{fig:DWSAlgorithm}.

To register calibration and distorted caustic patterns in the DWS, we utilize a custom multi-level Demon's algorithm to provide accurate non-rigid registration between the reference and distorted caustics \cite{Thirion1998}. A 5-level pyramid approach is utilized (scaling factor of 2 per level), with decreasing iterations per level to increase the speed of the algorithm. When complete, a precise x-y displacement map between the caustic patterns can be generated ($\boldsymbol{\vec{D}}(x,y)$), which, after scaling by the pixel size and dividing by the distance from the diffuser to the sensor ($f_D$), is approximately equivalent to the gradient of the transverse optical path difference induced by the introduction of refractive error:

\begin{equation}
\nabla \boldsymbol{\vec{\delta}}(x,y) \simeq  \frac{\boldsymbol{\vec{D}}(x,y)}{f_D}
\label{eq:DeltaDisplacement}
\end{equation}
\vskip 0.1in
The wavefront (W($\rho$, $\theta$)) can then be obtained by integration:

\begin{equation}
W(\rho, \theta) = \iint \nabla \boldsymbol{\vec{\delta}}(x,y) \,dx\,dy
\label{eq:DeltaIntegration}
\end{equation}
\vskip 0.05in
Note W($\rho$, $\theta$) has been converted to unit-circle normalized polar coordinates, and flipped about its vertical axis to correct for the reflection off the second beamsplitter, allowing direct comparison to the SHWS. The measured wavefront was decomposed into the Zernike basis (Equation~\ref{eq:ZernikeBasis}),

\begin{equation}
W(\rho, \theta) = \sum_{n = 1}^{N} c_i Z_i(\rho, \theta)
\label{eq:ZernikeBasis}
\end{equation}
\vskip 0.1in

\noindent and the fitted Zernike coefficients ($c_i$) were used to calculate the spherical and cylindrical error introduced by the trial lens, via Equations~\ref{eq:Axis}-\ref{eq:Phi2} below \cite{Dai2008,Schwiegerling}. 

\begin{equation}
\theta_A = \frac{1}{2} \arctan \frac{c_4}{c_6}
\label{eq:Axis}
\end{equation}

\begin{equation}
\phi_1 = -[\frac{2\sqrt{6}}{R_{max}^2}(c_4sin(2\theta_A)+c_6cos(2\theta_A))+\frac{4\sqrt{3}}{R_{max}^2}c_5] 
\label{eq:Phi1}
\end{equation}

\begin{equation}
\phi_2 = [\frac{2\sqrt{6}}{R_{max}^2}(c_4sin(2\theta_A)+c_6cos(2\theta_A))-\frac{4\sqrt{3}}{R_{max}^2}c_5] 
\label{eq:Phi2}
\end{equation}
\vskip 0.1in

Where $R_{max}$ represents the maximum pupil radius sampled by the wavefront sensor. An eyeglass prescription can be calculated as S = $\phi_1$, C = $\phi_2 - \phi_1$, and A = $\theta_A$, where S is the spherical refractive error power (D), C is the cylindrical refractive error power (D), and A is the cylindrical axis \cite{Schwiegerling}. Further, power vector notation is calculated from M = S + 0.5(C), $J_{0}$ = -0.5(C)cos(2$\theta_A$), and $J_{45}$ = -0.5(C)sin(2$\theta_A$) \cite{Thibos1997,Schwiegerling}.  The multi-level Demon's algorithm was implemented in C++ using ITK and run on an Intel Core i7 2.8 GHz CPU \cite{InsightConsortium2012}. The remainder of the DWS algorithm was implemented in MATLAB. All code, including image registration and Zernike fitting, is available upon request.

\subsection{Wavefront sensor performance Metrics}

The dynamic range, accuracy, and precision of refraction measurement of the DWS and the SHWS were evaluated over the [-24D, +24D] range of trial lenses tested for each illumination source utilized (LD, LD+LSR, and LED). The dynamic range was defined as the range of spherical refractive error over which each device produced a mean measurement within 0.25D of the correct value. Precision was defined as the mean of the standard deviation ($\overline{\sigma}$) of the five repeated spherical refractive error measurements for each system configuration across the dynamic range. Similarly, the accuracy of each device was calculated by the mean root-mean-square-error ($RMSE$) of the spherical refractive error measurements over the dynamic range. The accuracy and precision were also calculated for the DWS over a limited fine-value trial lens range determined by the SHWS DR for each illumination type ($RMSE_{FV}$ and $\overline{\sigma}_{FV}$). The theoretical number of resolvable prescriptions ($NRP_{th}$) was calculated as the dynamic range divided by theoretical sensitivity, outlined in Table~\ref{table:PredictedSensitivityDRNumPrescriptions}. The experimental NRP ($NRP_{exp}$) was calculated as the dynamic range divided by the $RMSE$. Measuring both of these metrics considers both a theoretical pixel-wise discretization of the NRP metric, as well as a functional definition, which allows for the sub-pixel accuracy of centroid fitting and caustic registration that occurs experimentally.

\section{Results and Discussion}

\subsection{Spherical error measurements} \label{ssec:DWSSHWSM}

\begin{figure}[h!]
\centering\includegraphics[width=12cm]{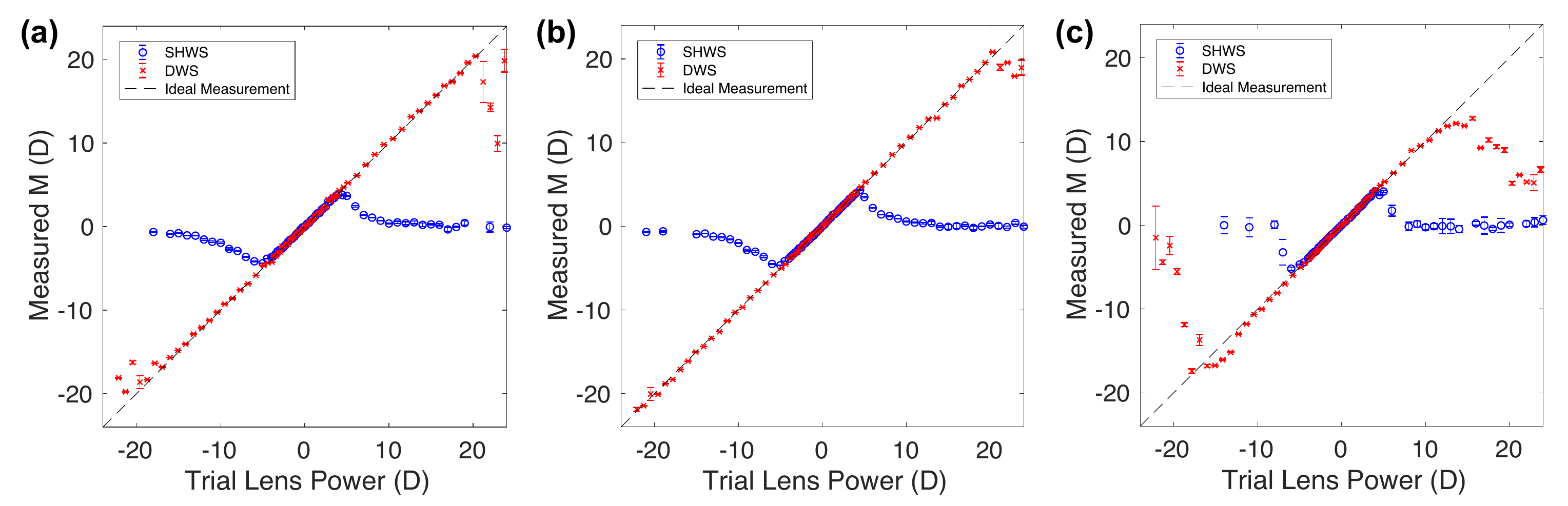}
\caption{M measurements and dynamic range results for (a) laser diode, (b) laser diode with laser speckle reducer, and (c) LED illumination for the DWS (red) and SHWS (blue) for [-24D, +24D] trial lens tested.}
\label{fig:DynamicRangeResults}
\end{figure}

In order to assess the dynamic range of the DWS and the SHWS, trial lenses ranging in spherical power from [-24D, +24D] were introduced adjacent to the model eye lens. The resulting measurements from the DWS and SHWS are plotted in Figure~\ref{fig:DynamicRangeResults} and the dynamic range, accuracy, sensitivity, and number of resolvable prescriptions were computed and displayed in Table~\ref{table:Measured_DR_MStd_RMSE} for each of the illumination types considered.

\begin{figure}[h!]
\centering\includegraphics[width=12cm]{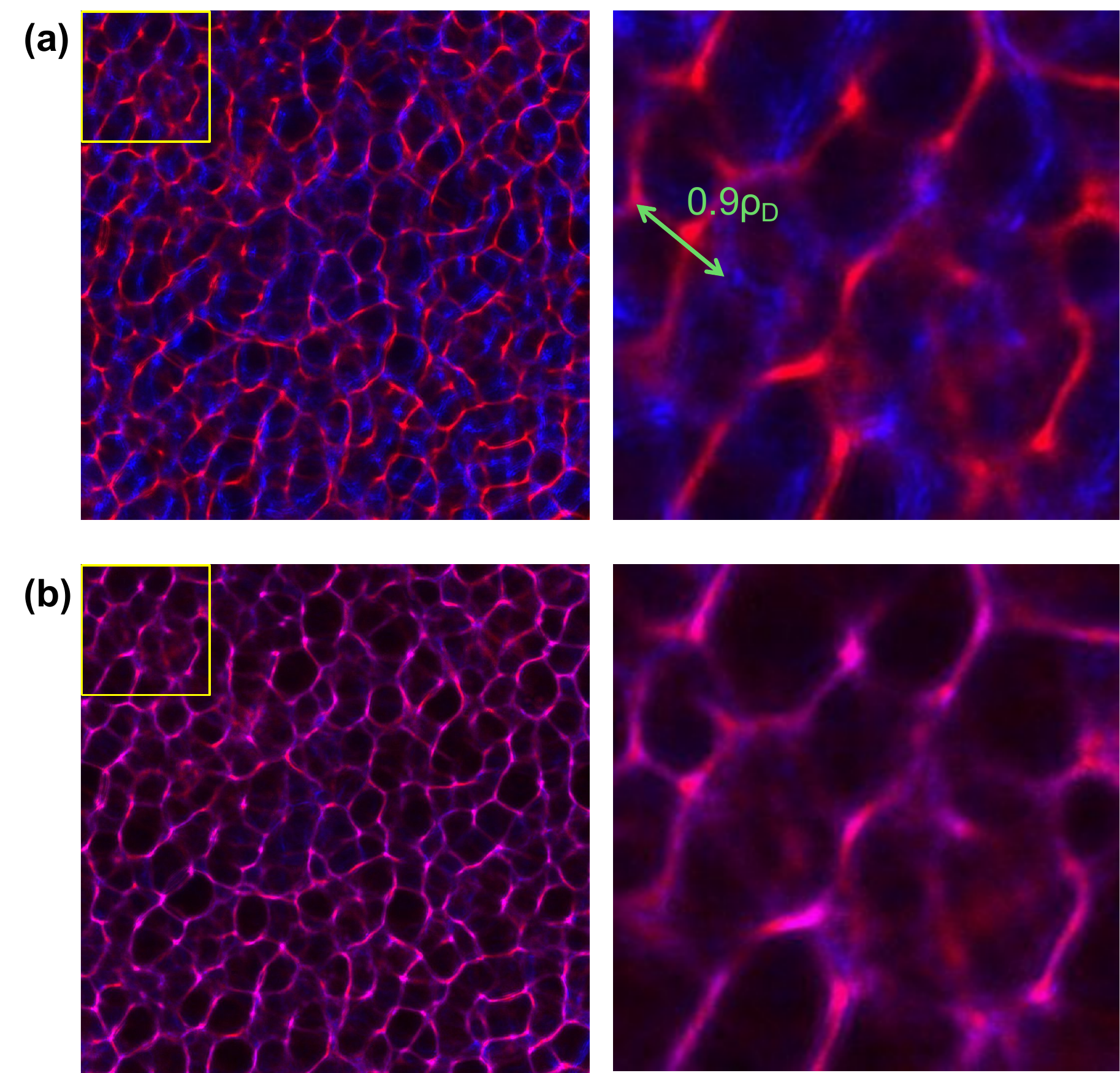}
\caption{Example caustic images (0D red, +16D blue) from the DWS before registration (a), and after registration (b). Full field of view (left) and outlined regions of interest (right). The green arrow shows an example registered feature exceeding the $\rho_{D}/2$ cutoff criteria that limits a conventional SHWS algorithm.}
\label{fig:DWSRegistrationExample}
\end{figure}

\captionsetup{labelfont=bf,textfont=bf}
\begin{table}[h!]
\caption{Spherical Error Measurement Results} \label{tab:Measured_DR_MStd_RMSE_title} 
\scriptsize
\centering
\begin{tabular}{c c c c c c c c c} 

 & \multicolumn{2}{c}{LD} && \multicolumn{2}{c}{LD+LSR} && \multicolumn{2}{c}{LED} \\
 \cline{2-3} \cline{5-6} \cline{8-9}
Measurement & SHWS & DWS && SHWS & DWS && SHWS & DWS \\ [0.5ex]
\hline
 DR (D) & [-3.75, +4.0] & [-17.0, +20.5] && [-4.0, +4.5] & [-22.0, +19.5] && [-4.5, +4.0] & [-10.5, +9.5] \\
 $\overline{\sigma}$ (D) & N/A & 0.027 && N/A & 0.009 && N/A & 0.005 \\
 $\overline{\sigma}_{FV}$ (D) & 0.022 & 0.024 && 0.019 & 0.004 && 0.024 & 0.005 \\
$RMSE$ (D) & N/A & 0.130 && N/A & 0.116 && N/A & 0.104 \\
$RMSE_{FV}$ (D) & 0.111 & 0.131 && 0.072 & 0.067 && 0.061 & 0.092 \\
 $NRP_{th}$ & 60 & 101  && 65 & 112 && 65 & 54 \\
 $NRP_{exp}$ & 70 & 288 && 118 & 358 && 139 & 192 \\
\hline
\end{tabular}
\label{table:Measured_DR_MStd_RMSE}
\end{table}
\captionsetup{labelfont=md,textfont=md}
\vskip 0.1in
Though the dynamic range of the SHWS with each illumination source is similar to predicted values in Table~\ref{table:PredictedSensitivityDRNumPrescriptions}, the dynamic range of the DWS is significantly greater than predicted values due to the $\rho/2$ cutoff criteria for the LD and LD+LSR illumination. The expanded dynamic range of the DWS can be attributed to the non-periodic nature of the caustic pattern being less susceptible to spot origin ambiguity as compared to the periodic spotfields in the SHWS. Figure~\ref{fig:DWSRegistrationExample} highlights the ability of the DWS to measure wavefronts from registered caustic displacements well beyond $\rho_D/2$. However, when the same Demon's image registration algorithm is applied to the SHWS spotfields, the registration begins to fail at a similar cutoff criteria to the conventional SHWS algorithm, further indicating it is the caustic pattern providing resistance to origin ambiguity rather than the DWS algorithm itself (see Appendix Section \ref{ssec:AppendixSHWSSpotfieldRegistration}). The dynamic range of the DWS is highest in the case of LD+LSR illumination, and the range of the instrument is relatively limited in the case of LED illumination. This is not due to a fundamental constraint imposed by using LED illumination, but rather due to limited optical power and a reduced signal-to-noise ratio provided by this source.

The data in Table~\ref{table:Measured_DR_MStd_RMSE} demonstrate that $NRP_{th}$ is significantly higher for the DWS than for the SHWS in all cases except the LED, and the $NRP_{exp}$ is significantly enhanced for the DWS for all illumination types. Though the SHWS out-performs the DWS in accuracy for every illumination type for each instrument's full respective dynamic range, the increased dynamic range of the DWS predominates to produce a significantly increased $NRP_{exp}$. The LD+LSR illumination offered the highest dynamic range and number of resolvable prescriptions. This is likely due to the relatively large power throughput and high signal-to-noise ratio allowed with the LD in combination with speckle noise reduction with the LSR. 

The differences in $RMSE$ between the SHWS and the DWS are likely due to the inherent predicted differences in sensitivity outlined in the Theory section in Table~\ref{table:PredictedSensitivityDRNumPrescriptions}, the greater error that comes with registering highly distorted caustics near the limit of the DWS dynamic range, and the effects of speckle noise on each algorithm. Despite having lower accuracy than the SHWS across its full dynamic range, the DWS is still able to produce measurements over its dynamic range with an $RMSE\leq$ 0.125D for LD+LSR and LED illumination, and is capable of providing accurate eyeglass prescriptions (typically prescribed in 0.25D increments). Additionally, the DWS demonstrates comparable $\overline{\sigma}$ with LD illumination, and improved repeatability with LD+LSR and LED illumination as compared to the SHWS over its full dynamic range, indicating the DWS algorithm has high test-retest repeatability. We note that for both the SHWS and the DWS, the measured sensitivity was better than the sensitivity predicted from a minimum 1-pixel displacement, indicating that both systems measure displacement with sub-pixel accuracy.

The DWS repeatability and accuracy was also calculated over the restricted dynamic range defined by the SHWS DR at each illumination type ($\overline{\sigma}_{FV}$ and $RMSE_{FV}$). In calculating precision with this restricted range, the DWS shows an improved precision in fine-value measurements for LD and LD+LSR illumination, while maintaining the same precision for LED illumination. It also demonstrates an increase in accuracy for fine-value measurements with LD+LSR and LED illumination, while performing similarly with LD illumination. Taken together, these results suggest that measurement of high Diopter refractive errors tends to cause more variability and error than fine-value measurements in speckle-reduced sources. This is expected considering the more distorted caustics should be more difficult to register due to larger displacement and intensity spreading. Interestingly, in measuring the fine-value range of refractive error dictated by the SHWS DR, we observe that the DWS with LD+LSR illumination outperforms the SHWS in both accuracy and repeatability.

The effects of speckle noise on measurement accuracy and precision are evident in both instruments. For the SHWS, the $RMSE$ was highest with laser diode illumination due to difficulty of centroid finding in the presence of speckle noise, a phenomena that has been previously reported \cite{Goncharov2015}. As expected, as the amount of speckle noise decreases with the introduction of the LSR, and further with the incoherent LED, the SHWS $RMSE$ decreases significantly. The $RMSE$ for the DWS also followed this trend, though to a lesser extent. This is due to the structure of the multi-level Demon's algorithm. When the registration algorithm initiates at the lowest level, the effects of speckle noise are suppressed by sampling, and as the algorithm progresses towards higher resolution images, it has already found a relatively accurate minima among the speckle noise.

\subsection{Cylindrical error measurements} \label{ssec:DWSSHWSJ0J45}

\begin{figure}[h!]
\centering\includegraphics[width=12cm]{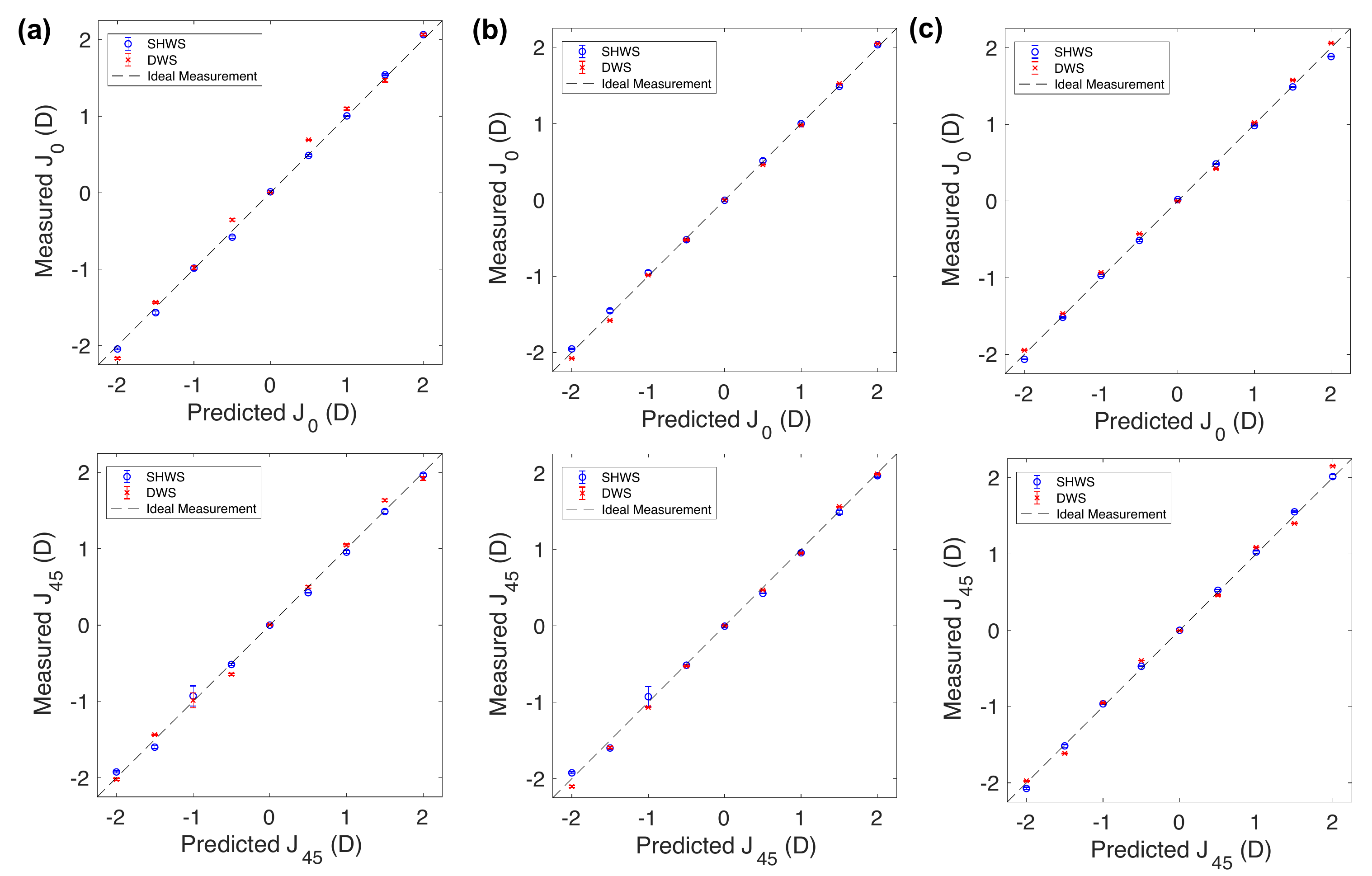}
\caption{$J_{0}$ (top row) and $J_{45}$ (botoom row) measurements for (a) laser diode, (b) laser diode with laser speckle reducer, and (c) LED illumination for the DWS and SHWS for [-4D, +4D] cylindrical trial lens tested at 0$^{\circ}$ and 45$^{\circ}$ axis.}
\label{fig:J0J45Results}
\end{figure}

To generate a complete eyeglass prescription, the sphero-cylindrical refractive error must be determined accurately. In this section, we compare the ability of the SHWS and DWS to refract a model eye with astigmatism by introducing cylindrical trial lenses over the range of [-4D, +4D] at angles of 0$^{\circ}$ and 45$^{\circ}$. This experiment tests a range of Jackson cross-cylinder values of $J_{0}$ and $J_{45}$ ranging from [-2D, +2D] (Figure~\ref{fig:J0J45Results}). Table~\ref{table:Measured_MStd_RMSE_Cyl} summarizes the results of these measurements.

\captionsetup{labelfont=bf,textfont=bf}
\begin{table}[h!]
\caption{Measured $J_{0}$ and $J_{45}$ mean standard deviation (\bm{$\overline{\sigma}$}) and root-mean-square-error (\bm{$RMSE$}) for the SHWS and DWS.} \label{tab:Measured_DR_MStd_RMSE_title} 
\centering
\begin{tabular}{c c c c}

\hline
Illumination & Measurement (D) & SHWS ($J_{0}$, $J_{45}$) & DWS ($J_{0}$, $J_{45}$) \\ [0.5ex]
\hline
LD & $\overline{\sigma}$ & (0.016, 0.027) & (0.014, 0.024) \\
 & $RMSE$ & (0.047, 0.058) & (0.108, 0.077) \\
\hline
LD+LSR & $\overline{\sigma}$ & (0.020, 0.029) & (0.003, 0.005) \\
 & $RMSE$ & (0.031, 0.058) & (0.045, 0.060) \\
\hline
LED & $\overline{\sigma}$ & (0.009, 0.013) & (0.004, 0.005) \\
 & $RMSE$ & (0.047, 0.036) & (0.058, 0.087) \\
\hline
\end{tabular}
\label{table:Measured_MStd_RMSE_Cyl}
\end{table}
\captionsetup{labelfont=md,textfont=md}

For all three illumination types and for both $J_{0}$ and $J_{45}$, the SHWS measured cylindrical powers with $< 0.06D$ $RMSE$. The DWS, while having slightly worse accuracy for the LD and LED illumination, is as accurate as the SHWS with the introduction of the laser speckle reducer. Similar to the M measurements before, the DWS demonstrates a lower $\overline{\sigma}$ and improved test-retest repeatability than the SHWS for LD+LSR and LED illumination, and comparable precision for LD illumination. From these results we can conclude that the DWS is capable of providing astigmatism measurement accurate enough for an eyeglass prescription.

\subsection{Optimized DWS for Autorefraction}

Based on the comparison of the SHWS and DWS under the three illumination types presented here, we can theorize what an ideal, optimized DWS setup would be for the field of autorefraction. We measured that the  0.5$^{\circ}$ holographic diffuser produces a sharp caustic pattern at $f_D$ = 5.15mm beyond the diffuser and has an effective diffuser pitch of $\rho_D$ = 338$\mu$m. The dynamic range of the diffuser utilized here was large and the sensitivity acceptable to provide eyeglass prescriptions when utilized with LD+LSR or LED illumination. However, similar to choosing lenslet arrays in order to trade off dynamic range for sensitivity in a SHWS, a different holographic diffuser could be utilized to increase the sensitivity of the DWS. Changing from the 0.5$^{\circ}$ holographic diffuser used here to a 0.2$^{\circ}$ holographic diffuser with similar $\rho_D$ should provide a longer $f_D$, an increased sensitivity comparable to the SHWS used here over its full dynamic range, and a similar improvement in number of resolvable prescriptions. Additionally, by assessing performance from sources with a range of speckle noise, we find that sources with limited speckle noise significantly improve DWS accuracy. The utilization of a laser diode with a laser speckle reducer provided the best balance of a large dynamic range and high accuracy. Further optimizing the system for higher power throughput and increased signal-to-noise ratio with LED illumination could also produce an inexpensive, accurate system with an improved dynamic range. If cost is not a primary concern in designing a DWS-based autorefractor, a super-luminescent diode would provide an excellent compromise of high power with low speckle noise. However, one of the advantages of using a holographic diffuser instead of a lenslet array is decreased cost - the diffuser used in this experiment is nearly 40x less expensive when comparing cost per clear aperture area. Lastly, despite the advantages of the DWS, the SHWS outperforms the DWS significantly with regards to speed. Though the data collection occurs in the same time frame and the memory requirements are nearly equivalent, the post-processing of the DWS data takes approximately 2 minutes per measurement, while the SHWS data processing occurs in a few seconds. In this report we did not optimize our registration algorithm for speed, and processing time could be significantly reduced by utilizing a GPU, an optical flow-based or other feature-based tracking algorithm, and by fitting the Zernike coefficients directly in the displacement domain. 

\section{Conclusions}

In this report we apply diffuser wavefront sensing to the field of autorefraction, demonstrating that a DWS is capable of providing accurate, high dynamic range refractive error measurements in a model eye by registering highly distorted caustics to reference images. We find that the non-periodic nature of the intensity distribution encoded by a holographic diffuser renders it resistant to spot origin ambiguity, which plagues conventional SHWS algorithms. Compared to a SHWS approach, we show that a DWS achieves significantly increased dynamic range and number of resolvable prescriptions. Additionally, the diffuser is nearly 40x less expensive per area than a lenslet array, making it ideal for affordable, large-range autorefraction devices. The implementation of this approach in modern portable autorefractors could lead to greater refractive care accessibility by enabling large-range wavefront sensing without any moving parts. Additionally, with recent advances in diffuser-based imaging and digital refocus technology, a diffuser-based approach may enable combined autorefraction and ocular imaging in the same device.

\section{Appendix} \label{sec:Appendix}

\subsection{Trial Lens Dynamic Range and Sensitivity Calculation} \label{ssec:AppendixDRSensCalc}

\begin{figure}[h!]
\centering\includegraphics[width=12cm]{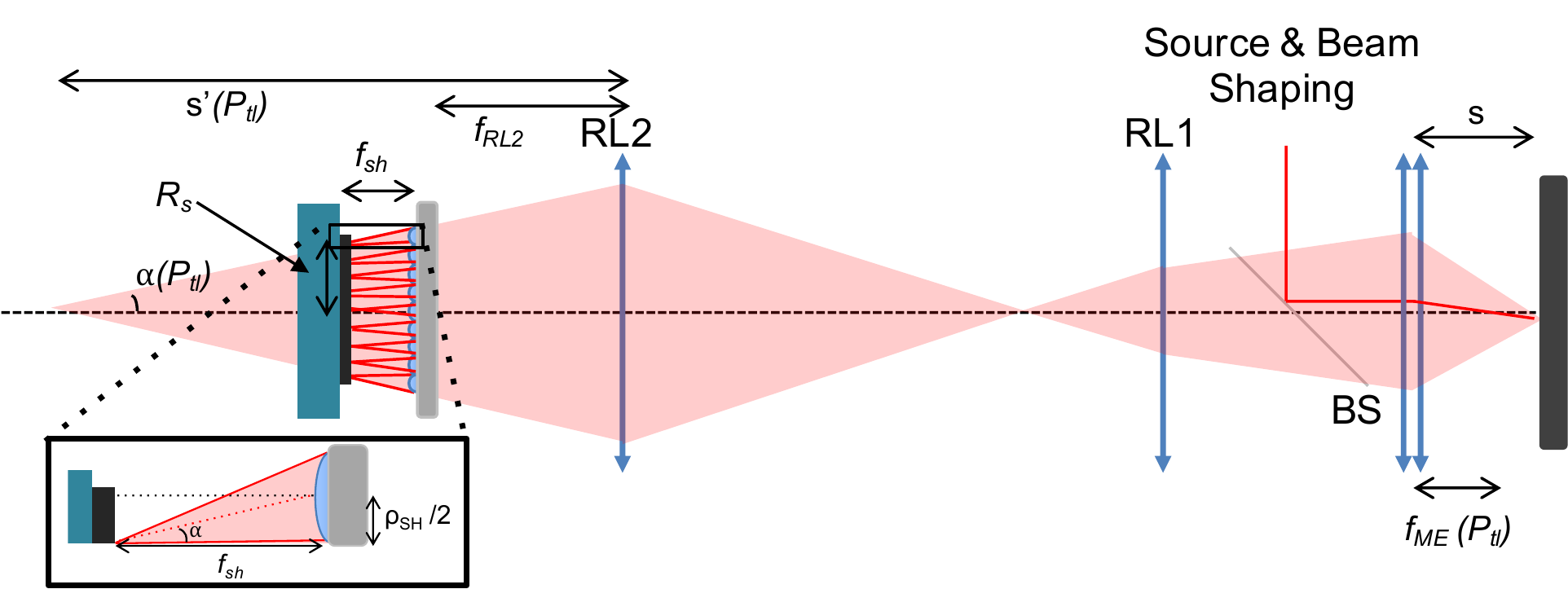}
\caption{Schematic of an example myopic refractive error measurement in the SHWS path. Note the thin lens approximation, similar right triangles defined by $\alpha$, and Equations~\ref{eq:SHWSMax}-\ref{eq:SHWSMin} allow a derivation of predicted dynamic range and sensitivity of the SHWS and DWS in terms of refractive error. The resulting inequality is shown below in Equation~\ref{eq:DRSensRatio}.}
\label{fig:DRSensDerivation}
\end{figure}

While the angular dynamic range and sensitivity of the SHWS can be calculated by Equations~\ref{eq:SHWSMax}-\ref{eq:SHWSMin}, the direct conversion of these angles to a myopic and hyperopic dynamic range and sensitivity is not as straightforward. To calculate these metrics, we model the system using a thin lens approximation, shown schematically for an example myopic refractive error in Figure~\ref{fig:DRSensDerivation}.

\begin{figure}[h!]
\centering\includegraphics[width=12cm]{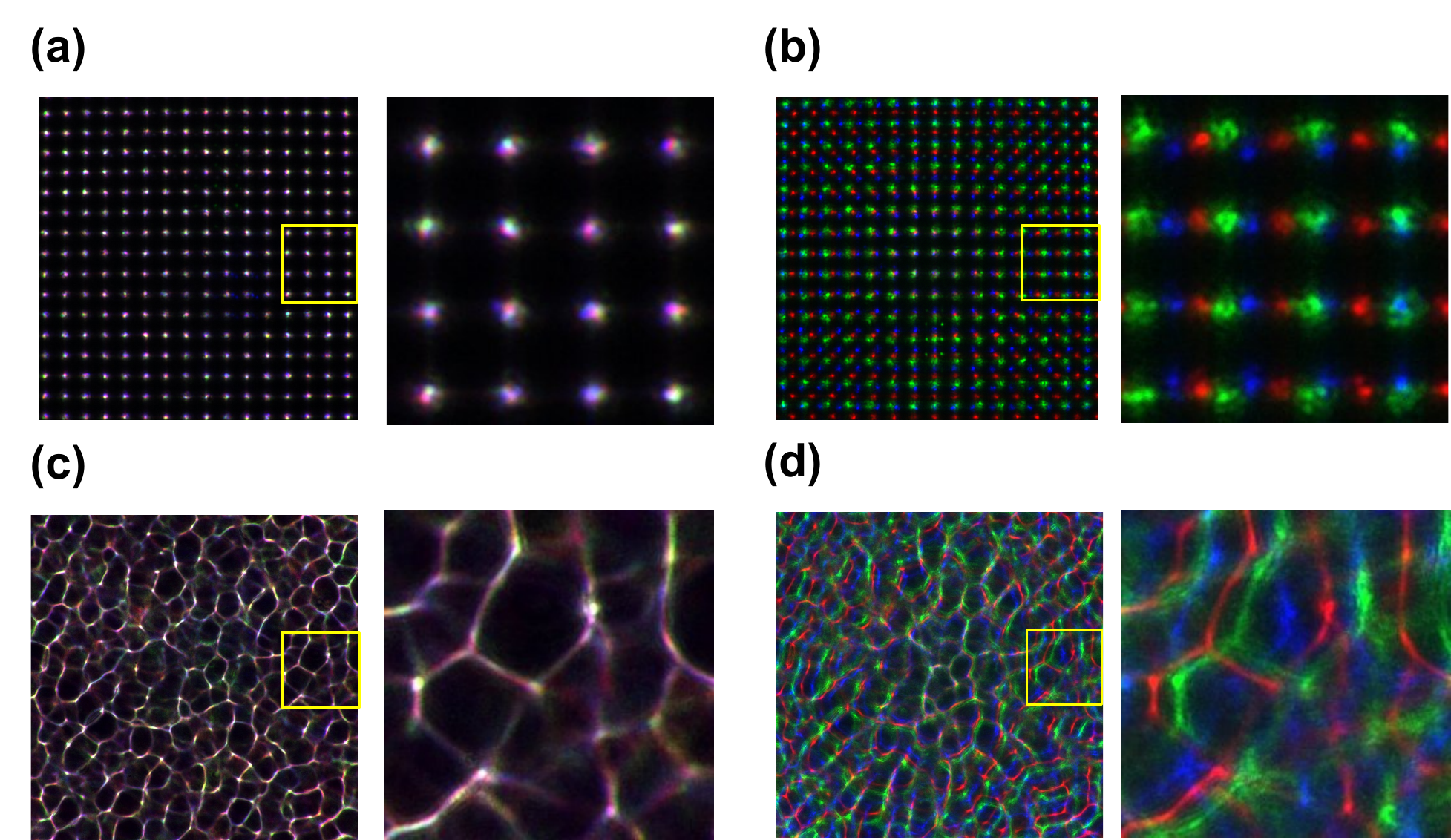}
\caption{Example spotfields (a)-(b) and caustics (c)-(d) demonstrating SHWS and DWS refractive error sensitivity and dynamic range predicted by Equation~\ref{eq:DRSensRatio}. For each image, red corresponds to an emmetropic eye, blue to myopic refractive error, and green to hyperopic refractive error. In each panel, a full frame 1024x1024 image is presented (left), with a 256x256 ROI outlined in yellow and magnified (right). (a) SHWS spotfield image showing sensitivity to $\pm$0.25D refractive error. Though spot intensity overlap occurs, centroid displacement is greater than $\Delta x$, visible by the color separation in the top right spot.  (b) SHWS spotfield image showing $\pm$4.5D refractive error near the limit of the instrument's dynamic range. Note spotfield displacement towards the periphery becomes ($\rho_{SH}$/2), visible where the green and blue spots overlap between two red spots. (c) DWS caustic images with $\pm$0.50D refractive error, near the predicted sensitivity of the instrument. Color separation is just barely visible towards the periphery of the image. (d) DWS caustics with $\pm$12D refractive error, near the predicted dynamic range of the instrument. Note towards the edge of the pupil, caustic displacement becomes nearly ($\rho_{D}$/2), visible where green and blue caustics overlap halfway between two red caustics.} 
\label{fig:DRSensSpotfieldsCaustics}
\end{figure}

While the distance from the model eye lens to the model retina stays fixed (s), the focal length of the model eye changes when a trial lens is introduced ($f_{ME}(P_{tl})$). For a myopic eye ($P_{TL}$ > 0D), the rays converge before the model retina, and thus $f_{ME}(P_{tl} > 0) < |s|$, while for a hyperopic eye ($P_{TL}$ < 0D), the rays converge behind the retina, and thus $f_{ME}(P_{tl} < 0) > |s|$. These two conditions result in a converging wavefront and real image, or diverging wavefront and virtual image, respectively. The final convergence point of the wavefront ($s'(P_{tl})$) after propagation through the model eye lens and the 4$f$ telescopic relay assembly can be calculated by applying multiple iterations of the thin lens equation, using the spot on the retina as the initial object. With this approach, using similar right triangles defined by $\alpha$ seen in Figure~\ref{fig:DRSensDerivation} and combining this with the SHWS angular sensitivity and dynamic range criteria of Equations~\ref{eq:SHWSMax}-\ref{eq:SHWSMin}, we derive the minimum and maximum measurable refractive error in terms of the wavefront sensor focal length ($f$), the pixel size ($\Delta x$), the sensor radius ($R_{S}$ = 2.381 mm and $R_{S}$ = 2.662 for the SHWS and DWS, respectively), and the pitch ($\rho$), shown below in Equation~\ref{eq:DRSensRatio}. A similar approach focusing on local wavefront criteria has been offered previously by Campbell \cite{Campbell2009}, but here we add a fixed $R_{S}$ dependency for the conventional SHWS algorithm applied in this report.

\begin{equation}
\Delta x < \frac{f}{|s'(P_{tl})-f_{RL2}-f|}*R_{S} < \frac{\rho}{2}
\label{eq:DRSensRatio}
\end{equation}
\vskip 0.1in
The middle term of the inequality is the maximum displacement experienced by a spot of the wavefront sensor at the periphery of the sensor, where $f$ refers to lenslet or diffuser focal length. As introduced with Equations~\ref{eq:SHWSMax}-\ref{eq:SHWSMin}, this displacement must be larger than $\Delta x$ and smaller than $\rho$/2 to be detectable within the aperture of a given lenslet. With these criteria, we calculate a predicted dynamic range of [-4.30D, +4.30D] and [-12.30D, +12.30D] and a predicted sensitivity of 0.13D and 0.37D, for the SHWS and DWS, respectively. Figure~\ref{fig:DRSensSpotfieldsCaustics} shows example caustic patterns and spotfield images at the predicted dynamic range and sensitivity extrema for each device, providing experimental verification of the validity of Equation~\ref{eq:DRSensRatio}.

\subsection{Effective Focal Length and Principal Plane Correction} \label{ssec:AppendixEFLBFLPP}

\begin{figure}[h!]
\centering\includegraphics[width=10cm]{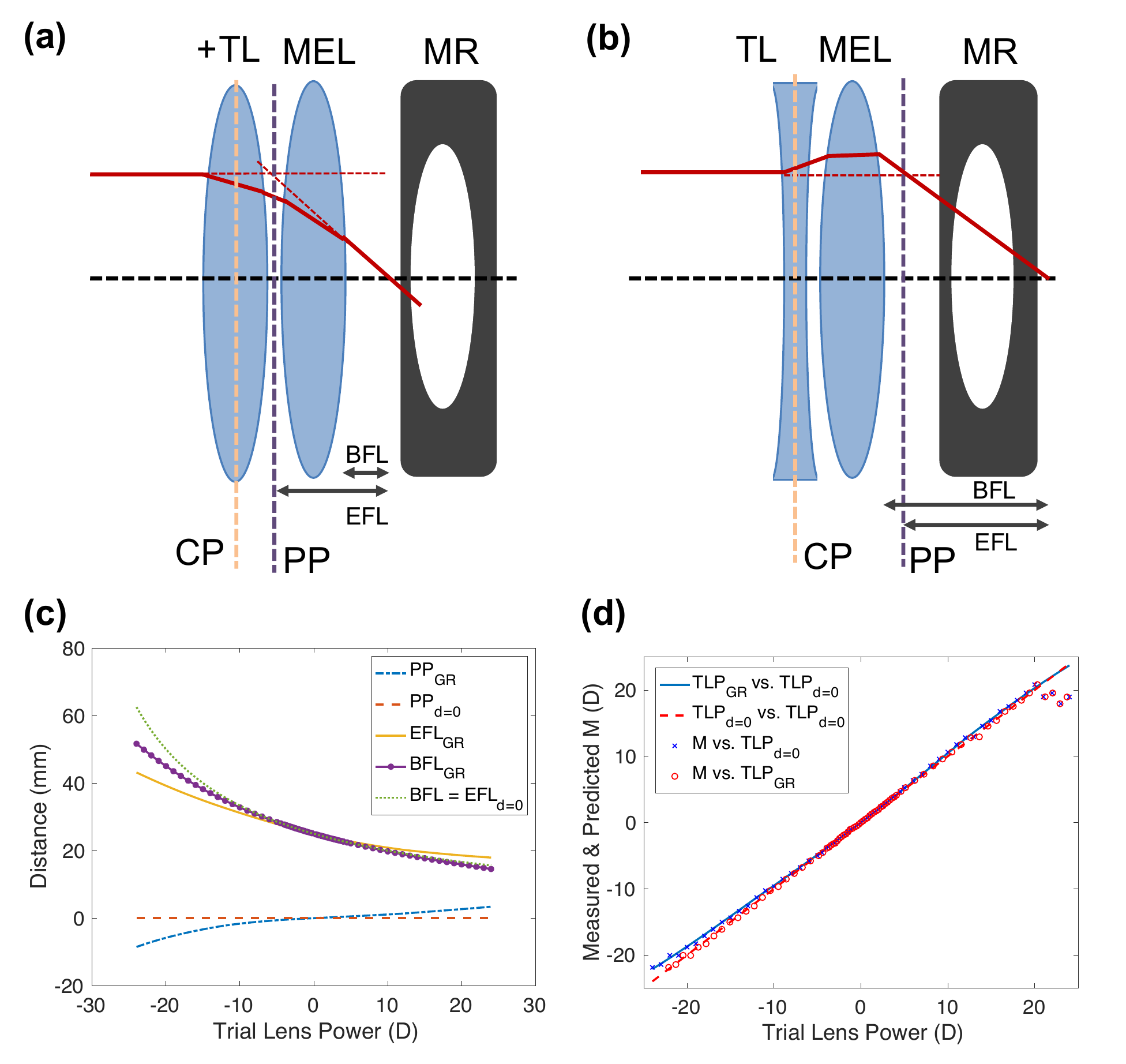}
\caption{(a) Schematic of the effective focal length (EFL), back focal length (BFL), principal plane (PP) location, and conjugate plane for the 1x, 4$f$ telescopic relay (CP) for $P_{TL}$ > 0D (myopic condition). (b) EFL, BFL, PP, and CP for $P_{TL}$ < 0D (hyperopic condition). Note that the distance between CP and PP is larger than in (a). (c) Calculation of changing EFL, BFL, and PP location for each of the [-24D, +24D] trial lens tested. Curves generated with Equation~\ref{eq:ThinLensMELPower} thin lens, 0mm separation approximation are denoted with "d = 0" subscript. Curves generated from Gaussian reduction and correction for moving principal plane are denoted with "GR" subscript. (d) Simulated trial lens power (TLP) measurements with thin lens d = 0mm approximation (red dashed line) and with Gaussian reduction correction (blue solid line). M measurements from DWS with LD+LSR illumination are plotted over these curves vs. $TLP_{d = 0}$ in blue x's before correction, and vs. $TLP_{GR}$ in red o's after correction.}
\label{fig:EFLBFLPP}
\end{figure}

To a first approximation, the power of the model eye lens with refractive error induced by adjacent placement of trial lenses can be well modeled with a thin lens, zero distance separation approximation:

\begin{equation}
P_{ME} = P_{MEL}+P_{TL}
\label{eq:ThinLensMELPower}
\end{equation}
\vskip 0.1in
Where $P_{ME}$ is the power of the model eye with refractive error, $P_{MEL}$ is the power of the emmetropic model eye lens, and $P_{TL}$ is the power of the trial lens(es) introduced. In practice, this approximation works well for small refractive error, but requires correction for larger refractive error. Here we model the back focal length (BFL), effective focal length (EFL), and principal plane (PP) location using a thin lens Gaussian reduction for each combination of trial lens used to better approximate the refractive error introduced. Figure~\ref{fig:EFLBFLPP}(a)-(b) provides a schematic of how these characteristics change with $P_{TL}$ > 0D and $P_{TL}$ < 0D trial lenses, the resulting calculations of EFL, BFL, and PP location (c), and the effect these have on the expected trial lens power measurement in the system (d).

In Figure~\ref{fig:EFLBFLPP}(d) we can see that the approximation of Equation~\ref{eq:ThinLensMELPower} begins to fail for high power trial lenses. The error is asymmetrically skewed towards the hyperopic range due to the alignment of the 1x, 4$f$ telescopic relay conjugate with the trial lens housing. While the trial lens is placed in front of the model eye lens, the principal plane is pushed behind the model eye lens for hyperopic refractive error, placed further from the conjugate plane of the telescopic relay (Figure~\ref{fig:EFLBFLPP}(b)). This introduces a larger error than the myopic condition, where the principal plane is pushed forward to be more closely aligned with the telescopic relay conjugate plane (Figure~\ref{fig:EFLBFLPP}(a)). The result of this correction is clearly demonstrated in Figure~\ref{fig:EFLBFLPP}(d), where the measured M DWS LD+LSR data is plotted vs. the simple approximation initially considered in Equation~\ref{eq:ThinLensMELPower} (blue x's), and vs. the predicted effective trial lens power when considering the changing EFL, BFL, and PP (red circles). Though this correction is important for validation of the DWS, we note that the change principal plane location for a human eye with refractive error will be significantly less than for the trial lenses considered here. Much of the error introduced is simply due to physical constraints of the housing of the trial lenses being placed adjacent to the model eye. In a real human eye with refractive error, the cornea and lens, are physically much closer together.

\subsection{Demon's Registration Applied to SHWS Spotfields} \label{ssec:AppendixSHWSSpotfieldRegistration}

\begin{figure}[h!]
\centering\includegraphics[width=10cm]{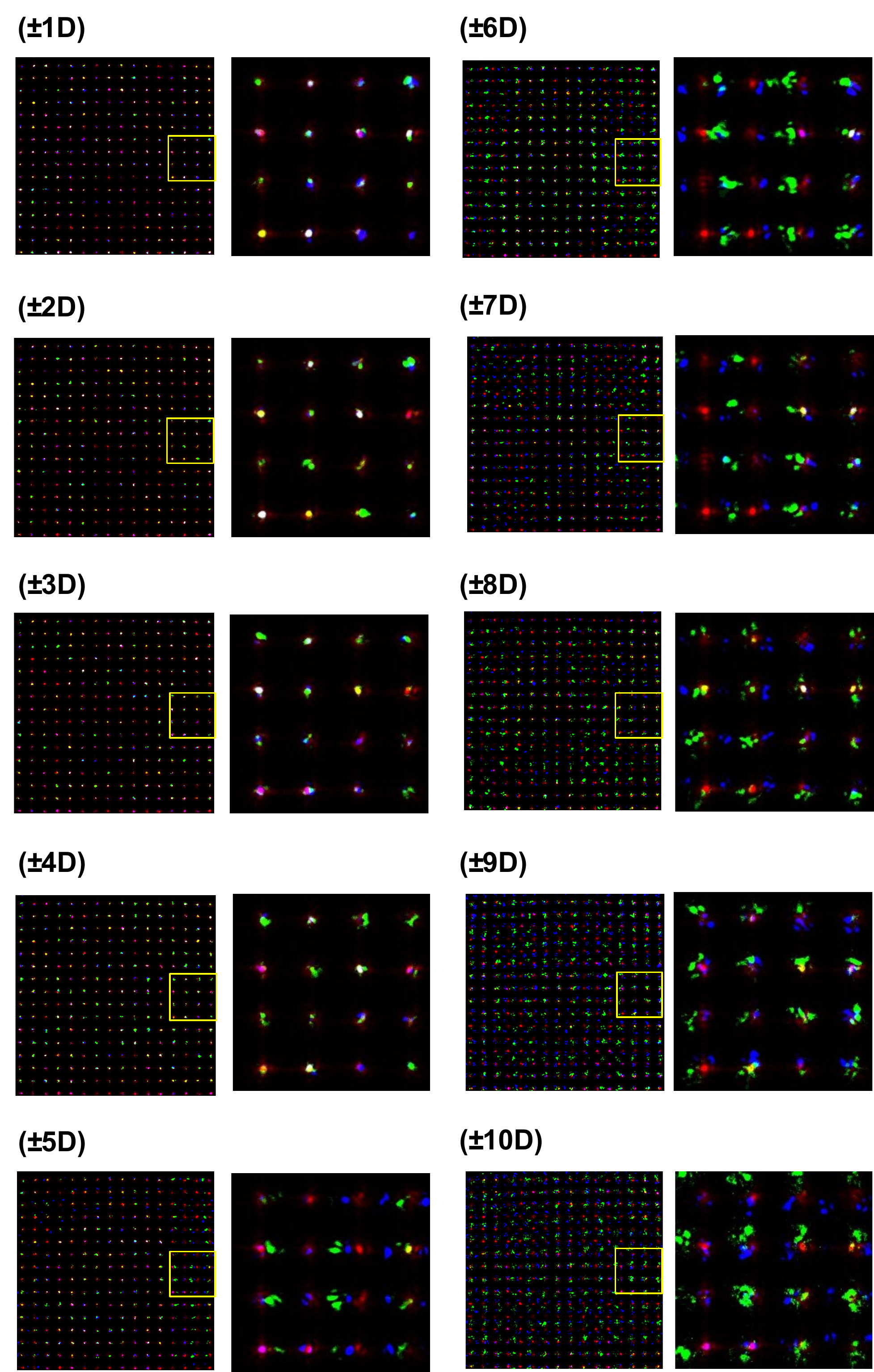}
\caption{Registration results of multi-level Demon's algorithm run on SHWS spotfield images from [-10D, +10D]. Each image includes the 0D emmetropic spotfield (red), the $P_{tl} > 0$ trial lens spotfield after registration (blue), and the $P_{tl} < 0$ trial lens after registration (green). Note that after $\pm$4D the registration algorithm begins to fail, which is approximately the same dynamic range found for the $\rho_{SH}$/2 criteria. This is in contrast to the DWS caustic intensity pattern, which is able to register beyond $\rho_{D}$/2, as seen in Figure~\ref{fig:DWSRegistrationExample}.}
\label{fig:SHWSSpotfieldRegistration}
\end{figure}

In order to further confirm that the non-periodic nature of the DWS caustics was the cause of the instrument's increased dynamic range, we applied the same multi-level Demon's non-rigid registration algorithm to the SHWS spotfield images from [-10D, +10D]. The results of this registration are shown in Figure~\ref{fig:SHWSSpotfieldRegistration}.

We see from this data that the multi-level Demon's algorithm applied to the SHWS spotfields begins to fail after $\pm$4D, indicating the periodic intensity distribution created by a lenslet array is prone to spot origin ambiguity. Application of the DWS algorithm to the SHWS spotfield data would yield approximately the same dynamic range as the conventional SWHS algorithm confining the spotfield displacement to the lenslet's field of view. Thus, we can be confident that it is truly the non-periodic nature of the caustic intensity distribution that affords the DWS resistance to origin ambiguity and an extended dynamic range (see Figure~\ref{fig:DWSRegistrationExample}).

\section*{Disclosures}

The authors are co-inventors on a provisional patent application assigned to Johns Hopkins University. They may be entitled to future royalties from intellectual property related to the technologies described in this article.


\begin{thebibliography}{10}
\newcommand{\enquote}[1]{``#1''}

\bibitem{Fricke2012}
T.~R. Fricke, B.~A. Holden, D.~A. Wilson, G.~Schlenther, K.~S. Naidoo,
  S.~Resnikoff, and K.~D. Frick, \enquote{{Global cost of correcting vision
  impairment from uncorrected refractive error},}
  {\protect\JournalTitle{Bulletin of the World Health Organization}}
  \textbf{90}, 728--738 (2012).

\bibitem{Durr2014}
N.~J. Durr, S.~R. Dave, E.~Lage, S.~Marcos, F.~Thorn, and D.~Lim,
  \enquote{{From Unseen to Seen: Tackling the Global Burden of Uncorrected
  Refractive Errors},} {\protect\JournalTitle{Annual Review of Biomedical
  Engineering.}} \textbf{16}, 131--153 (2014).

\bibitem{Naidoo2012}
K.~S. Naidoo and J.~Jaggernath, \enquote{{Uncorrected refractive errors},}
  {\protect\JournalTitle{Indian Journal of Ophthalmology}} \textbf{60},
  432--437 (2012).

\bibitem{Resnikoff2008}
S.~Resnikoff, D.~Pascolini, S.~P. Mariotti, and G.~P. Pokharel,
  \enquote{{Global magnitude of visual impairment caused by uncorrected
  refractive errors in 2004},} {\protect\JournalTitle{Bulletin of the World
  Health Organization}} \textbf{86 (1)}, 63--70 (2008).

\bibitem{Smith2009}
T.~S.~T. Smith, K.~D. Frick, B.~A. Holden, T.~R. Fricke, and K.~S. Naidoo,
  \enquote{{Potential lost productivity resulting from the global burden of
  uncorrected refractive error},} {\protect\JournalTitle{Bulletin of the World
  Health Organization}} \textbf{87}, 431--437 (2009).

\bibitem{Dandona2001}
R.~Dandona and L.~Dandona, \enquote{{Refractive error blindness},}
  {\protect\JournalTitle{Bulletin of the World Health Organization}}
  \textbf{79}, 237--243 (2001).

\bibitem{Resnikoff2012}
S.~Resnikoff, W.~Felch, T.-M. Gauthier, and B.~Spivey, \enquote{{The number of
  ophthalmologists in practice and training worldwide: a growing gap despite
  more than 200,000 practitioners},} {\protect\JournalTitle{British Journal of
  Ophthalmology}} \textbf{96}, 783--787 (2012).

\bibitem{WHO2010}
\enquote{{Action Plan for the prevention of avoidable blindness and visual
  impairment, 2009–2013},} Tech. rep. (2010).

\bibitem{Durr2015}
N.~J. Durr, S.~R. Dave, F.~A. Vera-Diaz, D.~Lim, C.~Dorronsoro, S.~Marcos,
  F.~Thorn, and E.~Lage, \enquote{{Design and Clinical Evaluation of a Handheld
  Wavefront Autorefractor},} {\protect\JournalTitle{Optometry and Vision
  Science}} \textbf{92}, 1140--1147 (2015).

\bibitem{Ciuffreda2015}
K.~J. Ciuffreda and M.~Rosenfield, \enquote{{Evaluation of the SVOne: A
  Handheld, Smartphone-Based Autorefractor},} {\protect\JournalTitle{Optometry
  and Vision Science}} \textbf{92}, 1133--1139 (2015).

\bibitem{Pamplona2010}
V.~F. Pamplona, A.~Mohan, M.~M. Oliveira, and R.~Raskar, \enquote{{NETRA :
  Interactive Display for Estimating Refractive Errors and Focal Range},}
  {\protect\JournalTitle{Association for Computing Machinery (ACM)}}  (2010).

\bibitem{Durr2018}
N.~J. Durr, S.~R. Dave, D.~Lim, S.~Joseph, T.~D. Ravilla, and E.~Lage,
  \enquote{{Quality of eyeglass prescriptions from a low-cost wavefront
  autorefractor evaluated in rural India: results of a 708-participant field
  study.}} {\protect\JournalTitle{bioRxiv}}  (2018).

\bibitem{Liang1994}
J.~Liang, B.~Grimm, S.~Goelz, and J.~F. Bille, \enquote{{Objective measurement
  of wave aberrations of the human eye with the use of a Hartmann-Shack
  wave-front sensor},} {\protect\JournalTitle{Journal of the Optical Society of
  America}} \textbf{11}, 1949--1957 (1994).

\bibitem{Applegate2014}
R.~Applegate, D.~Atchison, A.~Bradley, A.~Bruce, M.~Collins, J.~Marsack,
  S.~Read, L.~N. Thibos, and G.~Yoon, \enquote{{Wavefront Refraction and
  Correction},} {\protect\JournalTitle{Optometry and Vision Science}}
  \textbf{91}, 1154--1155 (2014).

\bibitem{Bruce2014}
A.~S. Bruce and L.~J. Catania, \enquote{{Clinical Applications of Wavefront
  Refraction},} {\protect\JournalTitle{Optometry and Vision Science}}
  \textbf{91}, 1278--1286 (2014).

\bibitem{Cheng2003}
X.~Cheng, N.~Himebaugh, P.~S. Kollbaum, L.~N. Thibos, and A.~Bradley,
  \enquote{{Validation of a Clinical Shack-Hartmann Aberrometer},}
  {\protect\JournalTitle{Optometry and Vision Science}} \textbf{80}, 587--595
  (2003).

\bibitem{Dai2008}
G.-M. Dai, \emph{{Wavefront Optics for Vision Correction}} (Bellingham, Walsh,
  2008).

\bibitem{Porter2006}
J.~Porter, H.~M. Queener, J.~E. Lin, K.~Thorn, and A.~Awwal, \emph{{Adaptive
  Optics for Vision Science. Principles, Practices, Design, and Applications}}
  (John Wiley {\&} Sons, Inc., 2006).

\bibitem{Platt2001}
B.~C. Platt and R.~Shack, \enquote{{History and principles of Shack-Hartmann
  wavefront sensing},} {\protect\JournalTitle{Journal of Refractive Surgery}}
  \textbf{17}, S573--S577 (2001).

\bibitem{Wittenberg1988}
S.~Wittenberg, \enquote{{The Badal Optometer Paradox},}
  {\protect\JournalTitle{American Academy of Optometry}} \textbf{65}, 285--291
  (1988).

\bibitem{Atchison1995}
D.~Atchison, A.~Bradley, L.~N. Thibos, and G.~Smith, \enquote{{Useful
  Variations of the Badal Optometer},} {\protect\JournalTitle{American Academy
  of Optometry}} \textbf{72}, 279--284 (1995).

\bibitem{Huang2103}
G.~Huang, H.~Jiang, K.~Matthews, and P.~Wilford, \enquote{{Lensless Imaging by
  Compressive Sensing},} {\protect\JournalTitle{IEEE ICIP}} pp. 2101--2105
  (2013).

\bibitem{Antipa2016}
N.~Antipa, S.~Necula, R.~Ng, and L.~Waller, \enquote{{Single-Shot
  Diffuser-Encoded Light Field Imaging},} in \emph{IEEE International
  Conference on Computational Photography,}  (IEEE, 2016).

\bibitem{Waller2018}
N.~Antipa, G.~Kuo, R.~Heckel, B.~Mildenhall, E.~Bostan, R.~Ng, and L.~Waller,
  \enquote{{DiffuserCam: lensless single-exposure 3D imaging},}
  {\protect\JournalTitle{Optica}} \textbf{5}, 1--9 (2018).

\bibitem{Engfei2017}
P.~Wu, Z.~Liang, X.~Zhao, L.~Su, and L.~Song, \enquote{{Lensless wide-field
  single-shot imaging through turbid media based on object-modulated
  speckles},} {\protect\JournalTitle{Applied Optics}} \textbf{56}, 3335--3341
  (2017).

\bibitem{Erto2017}
P.~Berto, H.~Rigneault, and M.~Guillon, \enquote{{Wavefront sensing with a thin
  diffuser},} {\protect\JournalTitle{Optics Letters}} \textbf{42}, 1--4 (2017).

\bibitem{Kane1988}
S.~Feng, C.~Kane, P.~A. Lee, and A.~D. Stone, \enquote{{Correlations and
  Fluctuations of Coherent Wave Transmission through Disordered Media},}
  {\protect\JournalTitle{Physical Review Letters}} \textbf{61}, 834--837
  (1988).

\bibitem{Freund1988}
I.~Freund, M.~Rosenbluh, and S.~Feng, \enquote{{Memory Effects in Propagation
  of Optical Waves through Disordered Media},} {\protect\JournalTitle{Physical
  Review Letters}} \textbf{61}, 2328--2332 (1988).

\bibitem{Gunjala2018}
G.~Gunjala, S.~Sherwin, A.~Shanker, and L.~Waller, \enquote{{Aberration
  recovery by imaging a weak diffuser},} {\protect\JournalTitle{Optics
  Express}} \textbf{26}, 21054 (2018).

\bibitem{Leroux2009a}
C.~Leroux and C.~Dainty, \enquote{{A simple and robust method to extend the
  dynamic range of an aberrometer},} {\protect\JournalTitle{Optics Express}}
  \textbf{17}, 19055--19061 (2009).

\bibitem{Pfund1998a}
J.~Pfund, N.~Lindlein, and J.~Schwider, \enquote{{Dynamic range expansion of a
  Shack–Hartmann sensor by use of a modified unwrapping algorithm},}
  {\protect\JournalTitle{Optics Letters}} \textbf{23}, 995--997 (1998).

\bibitem{Xia2010}
M.~Xia, C.~Li, L.~Hu, Z.~Cao, Q.~Mu, and L.~Xuan, \enquote{{Shack-Hartmann
  wavefront sensor with large dynamic range},} {\protect\JournalTitle{Journal
  of Biomedical Optics}} \textbf{15(2)}, 1--10 (2010).

\bibitem{Yu2014}
L.~Yu, M.~Xia, H.~Xie, L.~Xuan, and J.~Ma, \enquote{{Novel methods to improve
  the measurement accuracy and the dynamic range of Shack–Hartmann wavefront
  sensor},} {\protect\JournalTitle{Journal of Modern Optics}} \textbf{61}
  (2014).

\bibitem{Shinto2016}
H.~Shinto, Y.~Saita, and T.~Nomura, \enquote{{Shack–Hartmann wavefront sensor
  with large dynamic range by adaptive spot search method},}
  {\protect\JournalTitle{Applied Optics}} \textbf{55}, 5413 (2016).

\bibitem{Hongbin2008}
Y.~Hongbin, Z.~Guangya, C.~F. Siong, and L.~Feiwen, \enquote{{A tunable
  Shack–Hartmann wavefront sensor based on a liquid-filled microlens array},}
  {\protect\JournalTitle{Journal of Micromechanics and Microengineering}}
  \textbf{18}, 1--8 (2008).

\bibitem{Aita2015}
Y.~Saita, H.~Shinto, and T.~Nomura, \enquote{{Holographic Shack–Hartmann
  wavefront sensor based on the correlation peak displacement detection method
  for wavefront sensing with large dynamic range},}
  {\protect\JournalTitle{Optica}} \textbf{2}, 411--415 (2015).

\bibitem{Optotune2014}
Optotune, \enquote{{Laser speckle reduction with Optotune's laser speckle
  reducer LSR-3000 {\&} LSR-OEM},} {\protect\JournalTitle{Application Note}}
  pp. 1--14 (2014).

\bibitem{Schwiegerling}
J.~Schwiegerling, \emph{{Field Guide to Visual and Ophthalmic Optics}} (SPIE
  Press, 2004).

\bibitem{Thirion1998}
J.~P. Thirion, \enquote{{Image matching as a diffusion process: An analogy with
  Maxwell's demons},} {\protect\JournalTitle{Medical Image Analysis}}
  \textbf{2}, 243--260 (1998).

\bibitem{Thibos1997}
L.~N. Thibos, W.~Wheeler, and D.~Horner, \enquote{{Power Vectors: An
  Application of Fourier Analysis to the Descriptive and Statistical Analysis
  of Refractive Error},} {\protect\JournalTitle{Optometry and Vision Science}}
  \textbf{74}, 367--375 (1997).

\bibitem{InsightConsortium2012}
{Insight Consortium}, \enquote{{ITK - Segmentation {\&} Registration Toolkit},}
   (2012).

\bibitem{Goncharov2015}
A.~S. Goncharov, N.~G. Iroshnikov, A.~V. Larichev, and I.~P. Nikolaev,
  \enquote{{The impact of speckle on the measurement of eye aberrations},}
  {\protect\JournalTitle{Journal of Modern Optics}} \textbf{62}, 1775--1780
  (2015).

\bibitem{Campbell2009}
C.~E. Campbell, \enquote{{The range of local wavefront curvatures measurable
  with Shack-Hartmann wavefront sensors},} {\protect\JournalTitle{Clinical and
  Experimental Optometry}} \textbf{92}, 187--193 (2009).

\end{thebibliography}
\end{document}